\documentclass[a4paper]{jpconf}
\usepackage{graphicx}
\usepackage{cite}
\usepackage{amsmath}
\usepackage{amssymb}

\begin{document}
\title{Precise measurement of the \boldmath{$K_{e2} / K_{\mu 2}$} branching ratio and search for new physics beyond the Standard Model}

\author{S Bianchin (on behalf of the TREK/E36 Collaboration)}

\address{TRIUMF, 4004 Wesbrook Mall, Vancouver, BC V6T 2A3, CANADA}

\ead{s.bianchin@triumf.ca}

\begin{abstract}
\noindent The E36 experiment recently conducted at J-PARC by the TREK Collaboration will provide a precise mesurement of the decay ratio 
$ R_K = \Gamma(K^+ \rightarrow e^+\nu_e) / \Gamma(K^+ \rightarrow \mu^+\nu_{\mu}) $ with the aim of testing lepton universality, a basic assumption of the Standard Model (SM).
The SM prediction for the decay ratio $R_K$ is highly precise ($\delta R_K / R_K = 4 \times 10^{-4}$) and any observed deviation from this prediction would clearly indicate the existence of new physics.
The E36 experimental apparatus, designed to offer small systematic errors, was installed at the J-PARC K1.1BR kaon beam line in the fall of 2014, fully commissioned in the spring of 2015, and the production data were collected in the fall of 2015.
A scintillating fiber target was used to stop a beam of $1.2 \times 10^6 K^+$ per spill. 
The $K^+$ decay products were then momentum analyzed by a 12-sector superconducting toroidal spectrometer and charged particles were tracked with high precision by Multi-Wire Proportional Chambers (MWPC) combined with a photon calorimeter with a large solid angle (75\% of $4\pi$) and 3 different particle identification systems. Details of the E36 experimental apparatus, as well as the status of the data analysis will be presented.
\end{abstract}

%The $K^+$ decay products were then detected by a 12-sector superconducting toroidal spectrometer capable of tracking charged particles with high resolution, combined with a photon calorimeter with a large solid angle (75\% of $4\pi$) and 3 different particle identification systems. 
%Details of the E36 experimental apparatus, as well as the status of the data analysis and preliminary results will be presented.

\section{Introduction}
High precision electroweak measurements have been used extensively to test the Standard Model (SM) predictions. The SM successfully describes the results of most particle and nuclear physics experiments and it is believed that any deviation from its predictions would indicate the existence of new physics. Lepton universality being a key characteristics of the SM, the $K^{+} \rightarrow l^{+}\nu_{l}$ $(K_{l2})$ decay, which is the simplest leptonic decay among the $K^{+}$ decay channels, is therefore one of the best channels to perform such tests.
Any violation of lepton universality clearly indicates the existence of new physics beyond the SM.
By measuring the ratio of leptonic decay widths of the charged kaon, $R_K$, defined as

\begin{equation}
R_K = \frac{\Gamma(K^{+} \rightarrow e^+{\nu}_e)}{\Gamma(K^{+} \rightarrow {\mu}^+{\nu}_{\mu})},
\end{equation}

\noindent the $K_{l2}$ hadronic form factor cancels out and the sensitivity to new physics is greatly enhanced by the helicity suppression in $K_{e2}$. The SM prediction can be determined with very high precision (${\delta}R_K / R_K = 0.04\%$)~\cite{Cirigliano:2007}, giving a value of

\begin{equation}
R_K^{SM} = \frac{m_e^2}{m_{\mu}^2}\left(\frac{m_K^2 - m_e^2}{m_K^2 - m_{\mu}^2}\right)^2(1+{\delta}_r) = (2.477 \pm 0.001) \times 10^{-5},
\end{equation}

\noindent where $\delta_r$ is a radiative correction due the internal bremsstrahlung process (IB) which, by definition, contributes to $R_K$ (unlike the structure dependent process (SD)). A Minimal Super-Symmetric (SUSY) extension of the SM (MSSM) with R-parity conservation has recently been considered as a candidate for new physics to be tested by $R_K$~\cite{Masiero:2006,Masiero:2008,Girrbach}. In the case of $K_{l2}$, in addition to the $W^{\pm}$ exchange, a charged Higgs-mediated SUSY lepton flavor violating (LFV) contribution can be enhanced if accompanied by the emission of a $\tau$ neutrino. This effect can be described as

\begin{equation}
R_K^{LFV} = R_K^{SM} \left(1 + \frac{m_K^4}{M_{H^+}^4} \cdot \frac{m_{\tau}^2}{m_e^2} \Delta_{13}^2 \tan^6{\beta}\right)
\end{equation}

\noindent where $M_{H^+}$ is the mass of the charged Higgs and $\Delta_{13}$ ($\lesssim 10^{-3}$) is the term induced by the exchange of a Bino and a Slepton. Taking $\Delta_{13} = 5 \times 10^{-4}$, $\tan{\beta} = 40$ and $M_{H^+} = 500$ GeV, would yield to a value of $R_K^{LFV} = 1.013 \times R_K^{SM}$. It is therefore possible to reach a contribution at the percent level due to possible LFV enhancements arising in SUSY models.\\

\begin{figure}[!h]
\centering
\includegraphics[width=14cm]{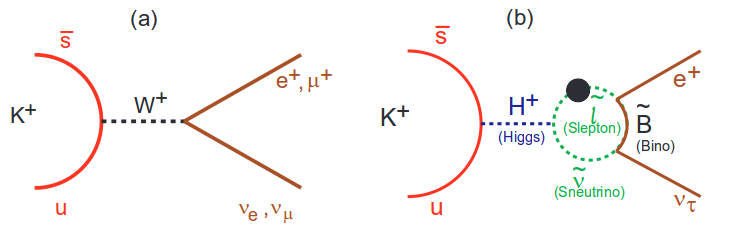}
\caption{Contributions to $R_K$ from (a) SM and (b) LFV SUSY. A charged Higgs-mediated LFV SUSY contribution can be strongly enhanced by the emission of a  $\tau$ neutrino.}
\end{figure}

\noindent Recent in-flight decay experiments, such as KLOE~\cite{Ambrosino:2009} and NA62~\cite{Lazzeroni:2013}, have measured the $R_K$ ratio leading to a current world average of $R_K = (2.488 \pm 0.010) \times 10^{-5}$.
The E36 experiment at the Japan Proton Accelerator Complex (J-PARC) by the TREK (Time Reversal Experiment with Kaons) collaboration aims to provide a competitive measurement of $R_K$ with different systematics.
%pursues the $R_K$ measurement using the stopped $K^+$ method which has completely different systematics from these previous in-flight-decay experiments.
%The E36 stopped $K^+$ decay experiment performed at the Japan Proton Accelerator Research Complex (J-PARC) by the TREK collaboration aims to improve by a factor of two the combined systematic and statistical uncertainty.\\

\section{\boldmath{$R_K$} measurements using stopped $K^+$ at J-PARC}
The E36 experiment, which is part of the TREK program at J-PARC, was set up and fully commissioned at the K1.1BR kaon beam line between fall 2014 and spring 2015. The data collection was completed by the end of 2015  using an upgraded version of the KEK-PS E246 12-sector superconducting toroidal spectrometer~\cite{MacDonald:2003} used in a previous T-violation experiment at KEK~\cite{Abe:1999,Abe:2004,Abe:2006}.
The incoming $K^+$ is tagged by the Fitch Cherenkov counter before stopping and decaying in the active target, which consists of 256 scintillating fibers ($3 \times 3 \times 200$~mm$^3$) oriented in the direction of the beam, providing a precise kaon stop location in the transverse plane.
Wrapped around the fiber target is a Spiral Fiber Tracker (SFT), which consists of two pairs of scintillating fiber ribbons of opposite helicity bundled together around the target, providing the longitudinal coordinate of the outgoing decay particles~\cite{Mineev:2015}.
This target+SFT system is surrounded by 12 time-of-flight counters (TOF1) and 12 aerogel~\cite{Tabata:2015} counters (AC) aligned with the 12 sectors of the toroidal spectrometer and forms the ``Central Detector''.\\
\noindent A highy segmented (768 crystals) large acceptance CsI(Tl) photon calorimeter barrel covering about 75\% of the total solid angle is used to identify and correct for structure dependent (SD) background events. The calorimeter features 12 holes (known as ``muon holes'') aligned with the sectors of the spectrometer allowing charged particles, such as $\pi^+$, $\mu^+$ and $e^+$, to be momentum analyzed by tracking with the Multi-Wire Proportional Chambers (MWPC) C2, C3 and C4.
At the exit of each magnet sector, another set of fast scintillator detectors (TTC and TOF2) and lead glass counters (PGC)~\cite{Miyazaki:2015} provides both trigger signals and $e^+/\mu^+$ particle identification.

\begin{figure}[!h]
\centering
\includegraphics[width=15cm]{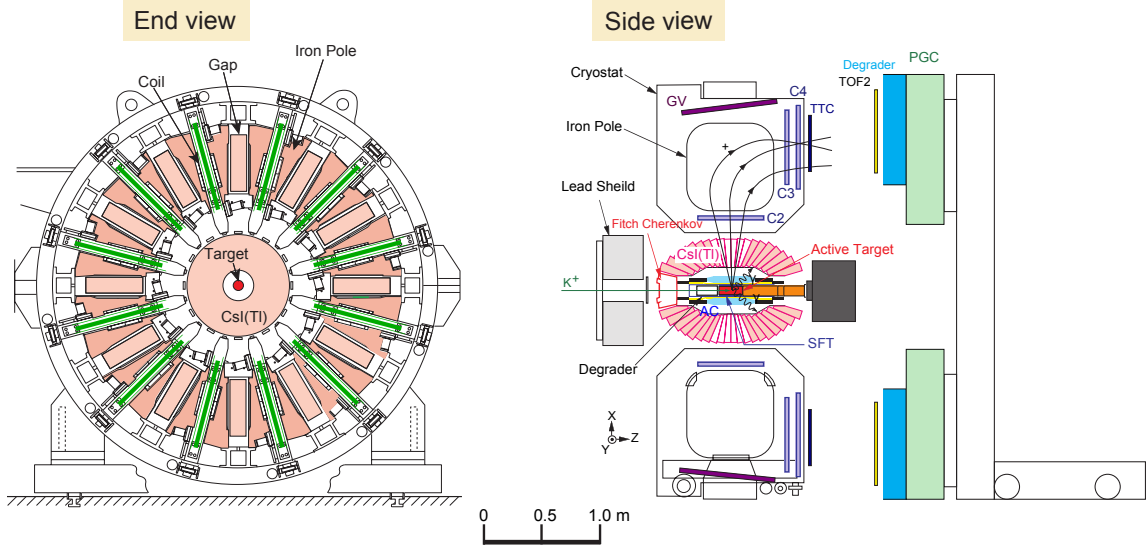}
\caption{Schematic end and side views of the E36 detector system at J-PARC.}
\end{figure}

\section{Status of analysis}
\noindent Three particle identification systems allow to redundantly distinguish between positrons, muons and pions. The threshold aerogel Cherenkov (AC) counters sensitive to positrons surround the target bundle, the time-of-flight (TOF) is measured between scintillators near the target (TOF1) and in each gap (TOF2), and lead glass counters (PGC) are located at the end of each gap to identify positrons by their shower.\\
\noindent The $e^+$ detection efficiency is estimated to be greater than 98\% for the aerogel (AC) and the lead glass (PGC) counters and the mis-identification probability to be about 3\% and 4\% respectively. The typical path length between TOF1 and TOF2 is about 250~cm and the time of flight is determined with a resolution of 200~ps; the time difference between $e^+$ and ${\mu}^+$ at $p \sim 236 - 247$~MeV/$c$ along this path being approximately 500~ps.\\
\noindent Figure 3 shows the particle momentum from tracking as a function of the ADC pulse height for AC (top left panel) and for PGC (top right panel).
The bottom left panel shows the momentum versus the squared mass from the time-of-flight analysis.
After applying threshold cuts in AC and PGC ADCs, the remaining events are shown in the bottom right panel where $K_{{\pi}2}$ is suppressed, $K_{{\mu}2}$ strongly reduced, and a clear separation between $K_{{\mu}2}$ and $K_{e3}$ events is observed. The expected $K_{e2}$ region is circled.\\
\noindent It is important to note that the analysis of the E36 experiment data is at an early stage and that these results are very preliminary.
Refinements of the particle identification method are being implemented and further improvements are expected.

\begin{figure}[!h]
\centering
\includegraphics[width=12cm]{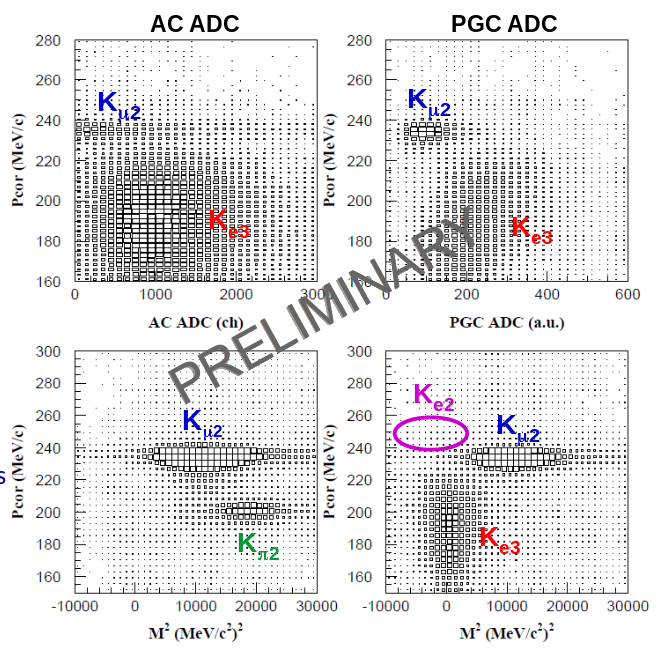}
\caption{Particle identification analysis. Top left: aerogel counter (AC) analysis (momentum vs ADC pulse height); top right: lead glass (PGC) analysis; bottom left: time-of-flight (TOF) analysis; bottom right: time-of-flight (TOF) analysis with threshold cuts on AC and PGC ADCs.}
\end{figure}

\section*{Acknowledgments}
\noindent This work is supported in part by NSERC and NRC in Canada, the KAKEN-HI Grants-in-Aid for Scientific Research in Japan, the DOE and NSF in the US, and in Russia by the Ministry of Science and Technology.

\end{document}